\documentclass[prd,twocolumn,showpacs,amsmath,amssymb,widetext,floatfix]{revtex4-1}
\usepackage{graphicx}
\usepackage{dcolumn}
\usepackage{bm}
\usepackage{epsfig}
\usepackage{color}
\usepackage{hyperref}

\hypersetup{
    colorlinks=true,
    linkcolor=red,
    citecolor=blue,
}

\def\la{\mathrel{\mathpalette\fun <}}
\def\ga{\mathrel{\mathpalette\fun >}}
\def\fun#1#2{\lower3.6pt\vbox{\baselineskip0pt\lineskip.9pt
  \ialign{$\mathsurround=0pt#1\hfil##\hfil$\crcr#2\crcr\sim\crcr}}}
\def\simgt{\mathrel{\lower0.6ex\hbox{$\buildrel {\textstyle >}
 \over {\scriptstyle \sim}$}}}
\def\simlt{\mathrel{\lower0.6ex\hbox{$\buildrel {\textstyle <}
 \over {\scriptstyle \sim}$}}}

\input epsf

\newcommand{\ompc}{{\rm Mpc}^{-1}}
\newcommand{\mpc}{{\rm Mpc}}
\newcommand{\mnras}{MNRAS}

\newcommand{\aj}{AJL}
\newcommand{\aap}{A$\&$A}

\def\be{\begin{equation}}
\def\ee{\end{equation}}
\def\ba{\begin{eqnarray}}
\def\ea{\end{eqnarray}}

\def\nn{\nonumber}

\newcommand{\eV}{{\,\rm eV}}
\newcommand{\bfs}{\mbox{\boldmath$s$}}
\newcommand{\bfr}{\mbox{\boldmath$r$}}
\newcommand{\bfx}{\mbox{\boldmath$x$}}
\newcommand{\bfk}{\mbox{\boldmath$k$}}
\newcommand{\bfv}{\mbox{\boldmath$v$}}
\newcommand{\bfu}{\mbox{\boldmath$u$}}
\newcommand{\bfp}{\mbox{\boldmath$p$}}

\newcommand{\jcap}{J. Cosmology Astropart. Phys.}

\begin{document}

\preprint{}

 \title{Measuring neutrino mass imprinted on the anisotropic galaxy clustering}
 
\author{Minji Oh$^{1,2}$, Yong-Seon Song$^{1,2}$}
\email{ysong@kasi.re.kr}
\affiliation{$^1$Korea Astronomy and Space Science Institute, Daejeon 34055, Korea} 
\affiliation{$^2$University of Science and Technology, Daejeon 34113, Korea}
\date{\today}

\begin{abstract}
The anisotropic galaxy clustering of large scale structure observed by the Baryon Oscillation Spectroscopic Survey Data Release 11 is analyzed to probe the sum of neutrino mass in the small $m_\nu\la 1\eV$ limit in which the early broadband shape determined before the last scattering surface is immune from the variation of $m_\nu$. The signature of $m_\nu$ is imprinted on the altered shape of the power spectrum at later epoch, which provides an opportunity to access the non--trivial $m_\nu$ through the measured anisotropic correlation function in redshift space (hereafter RSD instead of Redshift Space Distortion). The non--linear RSD corrections with massive neutrinos in the quasi linear regime are approximately estimated using one-loop order terms computed by tomographic linear solutions. We suggest a new approach to probe $m_\nu$ simultaneously with all other distance measures and coherent growth functions, exploiting this deformation of the early broadband shape of the spectrum at later epoch. If the origin of cosmic acceleration is unknown, $m_\nu$ is poorly determined after marginalising over all other observables. However, we find that the measured distances and coherent growth functions are minimally affected by the presence of mild neutrino mass. Although the standard model of cosmic acceleration is assumed to be the cosmological constant, the constraint on $m_\nu$ is little improved. Interestingly, the measured CMB distance to the last scattering surface sharply slices the degeneracy between the matter content and $m_\nu$, and the hidden $m_\nu$ is excavated to be $m_\nu=0.19^{+0.28}_{-0.17} \eV$ which is different from massless neutrino more than 68\% confidence.
\end{abstract}



\maketitle

\section{Introduction}

A neutrino is an elementary particle in the Standard Model (hereafter SM) of particle physics with three flavours whose existence was suggested as a ``desperate remedy" to explain the violation of energy conservation in nuclear beta decay~\cite{2007ConPh..48..195K}. The theoretically predicted particle was discovered~\cite{1956Sci...124..103C}, and became a fundamental building block of SM of particle physics. The detection of neutrino flavour oscillations also provides the indirect evidence of a non--trivial neutrino mass through the solar neutrino experiments~\cite{2004PhRvL..92r1301A, 2003PhRvL..90b1802E, 2005ConPh..46....1M}. The detection of possible neutrino mass constrains the lower bound on the sum of neutrino mass as $\sum m_{\nu}\ga 0.06\eV$ and $\sum m_{\nu}\ga 0.1\eV$ while assuming a normal mass hierarchy and an inverted hierarchy respectively~\cite{2016arXiv160503159C}. The tightest neutrino mass upper bound from laboratory has been reported to $\la 2\eV$~\cite{2002NuPhS.110..395B}. While this constraint bounded by particle physics experiments is expected to be improved by a few orders of magnitude by seeking for neutrinoless double beta decay events in the future~\cite{2002RvMP...74..663Z}, the more stringent upper neutrino mass bound is imposed through cosmological observations \cite{2006JCAP...10..014S, 2009ApJ...692.1060V, 2009A&A...500..657T, 2010JCAP...01..003R, 2010PhRvL.105c1301T, 2010MNRAS.406.1805M, 2010MNRAS.409.1100S, 2011ApJS..192...18K, 2014MNRAS.444.3501B, 2016PDU....13...77C}.

Alternatively, the neutrino mass can be probed by cosmological observations through the distinct clustering caused by the neutrino damping effect. The effect of massive neutrinos is imprinted on the recombination history through the alternated expansion history, which influences the shape of spectra determined at the last scattering surface. However, if $\sum m_{\nu}\la 1\eV$, the cosmic neutrinos become non--relativistic after the last scattering surface, and the transfer function with all massless neutrinos remains unchanged~\cite{2014PhRvD..89j3541S}. Light massive neutrinos are detectable by the Cosmic Microwave Background (hereafter CMB) experiment through the early Integrated Sachs--Wolfe (ISW) effect due to their being less relativistic around the last scattering surface and through the gravitational lensing effect developed at later epoch~\cite{2003PhRvL..91x1301K}. The Planck experiment constrains the neutrino mass upper bound as $\sum m_{\nu}\la 0.68\eV$ at 95\% confidence level~\cite{2015arXiv150201589P}.

The signature of neutrino mass is also imprinted on the large scale structure of the universe which is traced by galaxy distribution~\cite{1998PhRvL..80.5255H}. The observed galaxy clustering is plagued by uncertainties due to the non--linear mapping from the real to redshift spaces~\cite{Kaiser:1900zz, 1998ASSL..231..185H, 2012ApJ...748...78K, 2013PhRvD..88j3510Z}. This mapping is intrinsically non--Gaussian which leads to the infinite tower of cross--correlation pairs between density and velocity fields, and the non--perturbative suppression due to the randomness of infalling velocity is not theoretical predictable~\cite{2010PhRvD..82f3522T, 2016arXiv160300101Z}. Those all non--trivial corrections are not easily formulated with the cosmological model in which the neutrino becomes non--relativistic. But, if the targeted range of scale remains in the quasi linear regime, the known linear solutions can be fed tomographically at each epoch where the large scale structure evolves non--linearly. This approximation is proved to be applicable in our interesting scales~\cite{2016PhRvD..93f3515U}. 

Constraint on the neutrino mass is studied using the Baryon Oscillation Spectroscopic Survey Data Release 11 (hereafter BOSS DR 11) catalogue in this manuscript. Unlike the more conventional methodology to probe $m_\nu$ using galaxy clustering data, we suggest a new approach. The early broadband shape of the power spectrum determined at last scattering surface is altered when the neutrinos become non--relativistic~\cite{2011ARNPS..61...69W}. This shape departure is a unique signature of the non--trivial $m_\nu$, and it needs to be disclosed simultaneously with all other distance measures and coherent growth functions. When we apply the minimal theoretical prior for cosmic acceleration physics in which the structure evolves coherently after the last scattering surface, the $m_\nu$ constraint becomes weak and no confirming upper bound is discovered at small $m_\nu\la 1\eV$ limit. Although we impose the $\Lambda$CDM prior, the $m_\nu$ constraint little improves. With the $\Lambda$CDM prior, the CMB distance measure can be combined. This combination breaks the degeneracy between the matter content $\Omega_m$ and the neutrino mass $m_\nu$ significantly, and the massless limit of $m_\nu$ becomes distinguishable. The $m_\nu$ is measured to be $m_\nu=0.19^{+0.28}_{-0.17} \eV$ with the assumption of Gaussian probability distribution. We present the details in the following sections.

\section{Massive neutrino measurements}

\subsection{Large-scale structure with massive neutrino}

The cosmic neutrinos are decoupled when the expansion rate becomes dominant over the interaction rate with cosmic plasma. The neutrino is assumed to maintain Fermi--Dirac phase space distribution with decreasing temperature $T_{\nu}$ until the electron--positron annihilation. This does not correspond to the physical temperature after decoupling. If the neutrino decoupling is instantaneous, the temperature ratio to the photon temperature is given by, $T_{\nu}/T_{\gamma} = \left(4/11\right)^{1/3}$~\cite{2006PhR...429..307L}. In reality, this decoupling procedure is not instantaneous, and the annihilation occurs in the middle of the non--instantaneous decoupling procedure. This uncertainty is described by the non--trivial neutrino species number $N^{\mathrm{eff}}_{\nu}$ which is not the integer number 3. The exact value of $N^{\mathrm{eff}}_{\nu}$ was numerically computed to be 3.046, considering the non--instantaneous decoupling procedure as well as the quantum electrodynamic effect~\cite{2005NuPhB.729..221M}. This conventional $N^{\mathrm{eff}}_{\nu}$ is taken in this work, and treated to be constant, not a variable parameter. 

We probe the mild neutrino mass effect on galaxy clustering at later epoch, whilst the full recombination history is immune to this small neutrino mass. The cosmic neutrinos are relativistic before the last scattering surface, until then they freely stream and suppress the large scale structure formation. The early shape of spectrum is nearly identical to the shape with all massless neutrinos~\cite{2014PhRvD..89j3541S}. The neutrinos becomes non--relativistic after the recombination epoch, and those start to clump under the influence of local gravitational force. Suppression due to the free streaming effect is differently halted at various scales, and the broadband shape of spectrum is lately developed by this additional neutrino clustering. 

The early shape of spectra determined before the recombination epoch is influenced by initial conditions generated by inflation, and by the competition between radiative pressure resistance and gravitational infall during the radiation domination epoch \cite{Dodelson-Cosmology-2003}. When the fluctuations start to evolve coherently after matter-radiation equality, they have gone through a scale dependent shift from their own re--entering of the horizon to the recombination epoch. This early shape is nearly invariant with varying neutrino mass at $\sum m_{\nu}\la 1\eV$, and can be precisely computed by shape parameters of ($n_S$, $\omega_b$, $\omega_c$), which are determined by CMB experiments. The Planck experiment provides the constraints on the shape parameters with $n_S=0.97\pm 0.0060$, $\omega_b=0.022\pm0.00023$, and $\omega_c=0.12\pm 0.0022$ \cite{2015arXiv150201589P}.  

\begin{figure}
\begin{center}
\resizebox{3.2in}{!}{\includegraphics{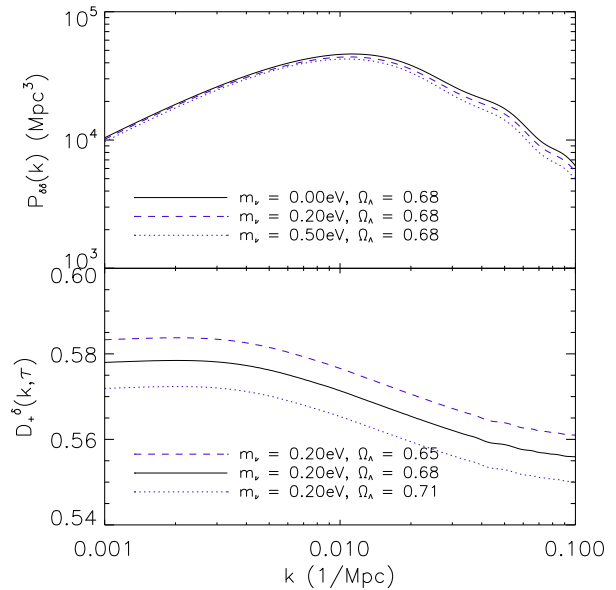}}
\end{center}
\caption{(Top panel) Linear power spectrum $P_{\delta\delta}(k)$ with one massive neutrino of mass 0.00 eV, 0.20eV, 0.50eV at redshift z=0.57. (Bottom) The growth function $D_{+}^{\delta}$ has coherently increasing amplitude as $\Omega_{\Lambda}$ decreases and scale-dependency which comes from non-zero neutrino mass, $m_{\nu}$ = 0.20 eV. See eq.8 for the definition of the scale-dependent growth function. The power spectra are computed using CAMB~\cite{Lewis:1999bs}.}\label{fig:Pklin}
\end{figure}

The small neutrino mass becomes non--relativistic at the late epoch, and the early broadband shape of spectra is altered by an additional neutrino element clumping to the existing dark matter clustering~\cite{2011ARNPS..61...69W}. In the small neutrino mass limit, one heavy neutrino mass dominates with normal hierarchy assumption, and two heavier neutrino masses are expected with inverted hierarchy assumption. The evolution of neutrino fluctuations develops differently with both assumptions despite the same $\sum m_{\nu}$. The distinction between different hierarchy assumptions is not significant under the sub--percentage change, while the shape of spectra is noticeably altered by the variation of the sum of neutrino masses. Hereby, the variation of $\sum m_{\nu}$ is represented by one heavy neutrino element mass variation. 

The geometric perturbations around the Friedman-Robertson-Walker universe are described by, 
\begin{equation}
ds^2 = -(1+ 2 \Psi) dt^2 + a(t)^2 (1 - 2 \Phi) \delta_{ij} dx^i dx^j.
\end{equation}
Those metric fluctuations are given by,
\begin{eqnarray}
- k^2\Psi = 4 \pi G a^2 \sum \rho_i\delta_i
\end{eqnarray}
where $i$ denotes the matter contents of baryon, cold dark matter or massive neutrino. The matter fluctuations $\delta_i$ are given by 

\begin{eqnarray}
\frac{\partial \delta_i}{\partial t} + (1+w_i)\left(\frac{\theta_i}{a} - 3\frac{\partial \Phi}{\partial t}\right) + 3H(c_{s,i} - w_i)\delta_i = 0, \nn
\label{continuity}
\end{eqnarray}
\begin{eqnarray}
\frac{\partial \theta_i}{\partial t} + H(1-3w_i)\theta_i &+& \frac{\partial w_i}{\partial t} \frac{\theta_i}{1+w_i}  \\ 
&+& \frac{c_{s,i}}{1+w_i}\frac{\nabla^2\delta_i}{a} - \frac{\nabla^2\sigma_i}{a} - \frac{\nabla^2\Psi}{a}=0 \nn \,,
\label{Euler}
\end{eqnarray}
where $w$, $c_s$ and $\sigma$ denote the equation of state, the sound speed and the anisotropic stress respectively. We will assume the irrotationality of fluid quantities and express the velocity field in terms of the velocity divergence $\theta_i = \nabla \cdot {\bf v}_i$. 

The scale dependent growth functions for $\delta$ and $\Theta = {\theta}/(aH)$ are denoted by $D_+^\delta$ and $D_+^\Theta$ respectively. We introduce the following parametrization of the growth rates
\ba
D_+^\delta(k, t) &=& G_{\delta}(t) F_\delta(k, t; m_{\nu}), \nn \\
D_+^\Theta(k, t) &=& G_{\Theta} (t) F_{\Theta}(k, t; m_{\nu}),
\label{eq:growth_rate}
\ea
where we defined $F_\delta(k, t; m_{\nu})$ and $F_{\Theta}(k, t; m_{\nu})$ so that $D_+^\delta(k, t) \to G_{\delta}(t)$ and $D_+^\Theta(k,t) \to G_{\Theta}(t)$ in the limit of $k \to 0$. In our small massive neutrino analysis, the early broadband shape with BAO signature is imprinted before the last scattering surface in which all neutrinos behave relativistic. When this early BAO physics is probed by CMB experiments, the baryon and cold dark matter contents are determined regardless of neutrino mass at $m_\nu\la1\eV$. In the top panel of Fig.~\ref{fig:Pklin}, the power spectra with varying neutrino masses of $m_\nu=(0.0,0.2,0.5)\eV$ are presented with the fixed $\Omega_\Lambda=0.68$. The neutrino becomes non--relativistic at $z \la 1000$ with $m_\nu=(0.0,0.2,0.5)\eV$. The shapes of those spectra are the same until $z \sim 1000$, and the distinct late time broadband shapes are developed due to different epochs when neutrinos become non--relativistic~\cite{2004JCAP...12..005R, 1983ApJ...274..443B, 1996ApJ...471...13M}. This departure from the early shape with $m_\nu=(0.2,0.5)\eV$ is presented by dash and dotted curves respectively. When $\Omega_\Lambda$ and $\Omega_m$ are fixed with varying $m_\nu$, the spectra with different $m_\nu=(0.0,0.2,0.5)\eV$ become the same at $k\to 0$ limit. However, the Hubble constant becomes slightly bigger with increasing $m_\nu$, while $\Omega_\Lambda$ is fixed. The $P_{\delta\delta}(k)$ is presented in $\mpc^3$ unit in which BAO peaks are invariant with varying $m_\nu$ and the overall amplitude is slightly decreases by $h^{-3}$. 


In the bottom panel of Fig.~\ref{fig:Pklin}, $D_+^\delta(k, t)$ is presented with the fixed $F_\delta(k, t; m_{\nu})$ at $m_{\nu}=0.2\eV$, while $\Omega_\Lambda$ varies. The dash, solid and dotted curves represent $D_+^\delta(k, t)$ with $\Omega_\Lambda=(0.65,0.68,0.71)$ respectively. The coherent shift of $D_+^\delta(k, t)$ presented in the figure verifies that the effect on structure formation by varying dark energy model can be represented by $G_{\delta}$ and $G_{\Theta}$ parameters. We use $G_b$ which represents the combined variation of $b$ and $G_\delta$, as both are not clearly separated.

\subsection{RSD with massive neutrino}

In practice, the inhomogeneous density traced by galaxy distribution is small perturbation to the homogeneous background observed in redshift space. The mapping formulation from real to redshift spaces causes the non--perturbative effect to the power spectrum even at linear regime in which the leading order terms of perturbations dominate \cite{Kaiser:1900zz, 1998ASSL..231..185H, 2012ApJ...748...78K, 2013PhRvD..88j3510Z}. The observed perturbations are altered by the peculiar motion of tracers along the line of sight, $\bfs=\bfr+\bfv \cdot \hat{z}/aH$, where $\bfr$ and $\bfs$ denote vector distances in real and redshift spaces respectively. Then the observed power spectrum is given by \cite{2010PhRvD..82f3522T},
\begin{equation}
P^{\rm(S)}(k,\mu)=\int d^3\bfx\,e^{i\,\bfk\cdot\bfx}
\bigl\langle e^{j_1A_1}A_2A_3\bigr\rangle\,, 
\label{eq:Pkred_exact}
\end{equation}
in which we define
\begin{eqnarray}
&j_1= -i\,k\mu ,\nonumber\\
&A_1=u_z(\bfr)-u_z(\bfr'),\nonumber\\
&A_2=\delta(\bfr)+\,\nabla_zu_z(\bfr),\nonumber\\
&A_3=\delta(\bfr')+\,\nabla_zu_z(\bfr'),\nonumber
\end{eqnarray}
where $\bfx=\bfr-\bfr'$, $\bfu\equiv-\bfv/(aH)$, $u_z$ is the velocity component along the line of sight. We consider the targeted galaxies from BOSS DR11 in the deep space in which the plane parallel approximation can be adopted, then $\mu$ denotes the cosine of the angle between $\bfk$ and the line of sight.

The effect of RSD mapping appears as non--Gaussian contribution in Eq.~\ref{eq:Pkred_exact}. The full expansion of Eq.~\ref{eq:Pkred_exact} contains the non--perturbative function caused by the randomness of infalling velocities which is known as the Finger--of--God effect (hereafter FoG) \cite{1972MNRAS.156P...1J}, and the higher order polynomials due to the non--trivial cross--correlation between density and velocity fields. It is given by \cite{2012PhRvD..86j3528T, 2016arXiv160300101Z},
\ba
\label{eq:Pkred_final}
P^{\rm (S)}(k,\mu)&=&D^{\rm FoG}(k\mu\sigma_z)[P_{\delta\delta}+2\mu^2P_{\delta\Theta}+\mu^4P_{\Theta\Theta}  \\
&&+A(k,\mu)+B(k,\mu)+T(k,\mu)+F(k,\mu)] \nonumber.
\ea
The $D^{\rm FoG}$ and $F(k,\mu)$ represent the one--point and the correlated FoG's respectively, which are given by~\cite{2016arXiv160300101Z},
\ba
D^{\rm FoG}_{\rm 1pt}(k\mu)&=&\exp\left\{j_1^2\sigma_z^2+2\sum_{n=2}^{\infty}j_1^{2n}\sigma_z^{2n}\frac{K_{2n}}{(2n)!}\right\}, \\
F(k,\mu)&=& -j_1^2\,\int d^3\bfx \,\,e^{i\bfk\cdot\bfx}\,\,\langle u_z u_z'\rangle_c\langle A_2A_3\rangle_c\,.
\ea
The one--point FoG effect denoted by $D^{\rm FoG}$ is dominated by the first order term with negligible $K_{2n}$ contributions. The $\sigma_z$ represents one--point velocity dispersion along the line--of--sight. There is no trustable theoretical model to predict $\sigma_z$, and it is set to be the scale independent free parameter which is simultaneously fitted with all other cosmological observables. The $F(k,\mu)$ represents the leading order correlated FoG effect. The higher order polynomials due to density--velocity cross--correlation are given by,
\begin{eqnarray}
  A(k,\mu)&=& j_1\,\int d^3\bfx \,\,e^{i\bfk\cdot\bfx}\,\,\langle A_1A_2A_3\rangle_c,\nonumber\\
  B(k,\mu)&=& j_1^2\,\int d^3\bfx \,\,e^{i\bfk\cdot\bfx}\,\,\langle A_1A_2\rangle_c\,\langle A_1A_3\rangle_c,\nonumber\\
  T(k,\mu)&=& \frac{1}{2} j_1^2\,\int d^3\bfx \,\,e^{i\bfk\cdot\bfx}\,\,\langle A_1^2A_2A_3\rangle_c,\nonumber.
\end{eqnarray}
In this manuscript, we consider low resolution experiment at linear regime in which the cancellation of $F+T$ combination leaves non--significant residuals, and theoretical models are effectively reduced to the following form \cite{2010PhRvD..82f3522T},
\ba
\label{eq:Pkth}
P^{\rm (S)}(k,\mu)&=&e^{-(k\mu\sigma_z)^2}\left[P_{gg}+2\mu^2P_{g\Theta}+\mu^4P_{\Theta\Theta} +A+B\right].\nn\\
\ea
where the subscript $g$ denotes the perturbed galaxy distribution, and the linear coherent bias formulation is adopted, $\delta_g=b\delta$ \cite{2013MNRAS.432..743N}.

The non--linear gravitational evolution with massive neutrino is computed using the resummed perturbation theory called {\tt RegPT} \cite{2008ApJ...674..617T, 2012PhRvD..86j3528T}, in which an ill-behaved expansion leading to unwanted UV behaviour is removed by suppression factor at small scales. We compute the non--linear corrections up to one loop by feeding the linear power spectrum at each redshift which is the solution of the full linear Boltzmann equation. There has been a few alternative approaches developed to improve theoretical prediction, but it was confirmed that TNS method provides more precise description for matter fluctuations with massive neutrino at quasi linear regime, exploiting the simulations at $z=0.5$ \cite{2016PhRvD..93f3515U}. In this manuscript, we adopt this study.
The loop correction terms $A(k,\mu)$ and $B(k,\mu)$ are similarly derived using the following formulation \cite{2013PhRvD..87h3509T}, 
\ba
A&=&b^3\sum_{n=1}^3\sum_{a,b=1}^2\mu^{2n}\left(\frac{G_{\Theta}}{G_{b}}\right)^{a+b-1}
\frac{k^3}{(2\pi)^2}\int_0^\infty dr\int_{-1}^1 dx
\nonumber\\
&&\times\Bigl\{
A^n_{ab}(r,x)\,B_{2ab}(\bfp,\bfk-\bfp,-\bfk)
\nonumber\\
&&\qquad+\widetilde{A}^n_{ab}(r,x)B_{2ab}(\bfk-\bfp,\bfp,-\bfk)\Bigr\},
\label{eq:Aterm}
\\
B&=&b^4\sum_{n=1}^4\sum_{a,b=1}^2\mu^{2n}\left(-\frac{G_{\Theta}}{G_{b}}\right)^{a+b}
\frac{k^3}{(2\pi)^2}\int_0^\infty dr\int_{-1}^1 dx
\nonumber\\
&&\times
\,B^n_{ab}(r,x)\frac{P_{a2}(k\sqrt{1+r^2-2rx})P_{b2}(kr)}{(1+r^2-2rx)^a}.
\label{eq:Bterm}
\ea
where $r=p/k$, $x=\bfk\cdot\bfp/(k\,p)$ and $G_{b}=bG_{\delta}$. Here, the 
functions $P_{ab}$ and $B_{abc}$ are the power spectrum and bispectrum of the two-component multiplet $\Psi_a=\left(\delta,-\frac{\Theta}{f^2}\right)$. The non-vanishing coefficients, $A^n_{ab}$, $\widetilde{A}^n_{ab}$, and $B^n_{ab}$,  are those presented in Sec.~III-B2 of Ref.~\cite{2013PhRvD..87h3509T} and Appendix A of Ref.~\cite{2010PhRvD..82f3522T}, respectively.  

\begin{figure}
\begin{center}
\resizebox{3.3in}{!}{\includegraphics{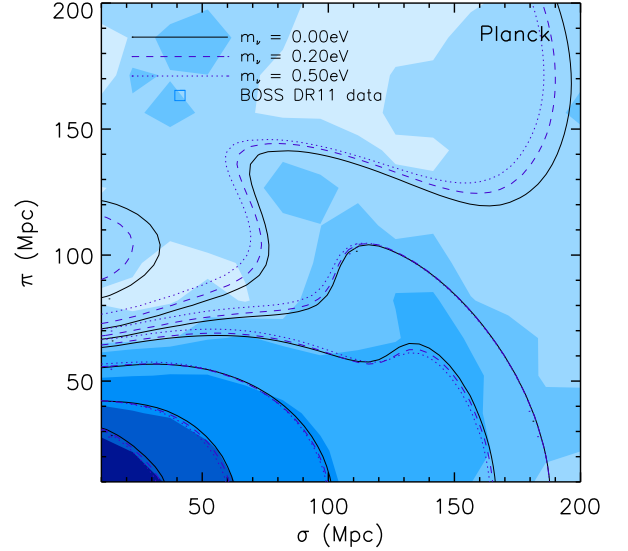}}
\end{center}
\caption{The measured two--point correlation function $\xi(\sigma,\pi)$ is presented as the filled blue contour, and the theoretical predictions of $\xi(\sigma,\pi)$ are shown as the unfilled solid, dash and dotted black contours with $m_\nu=0,0.2,0.5\eV$ respectively. The contour levels are $\xi(\sigma,\pi)=(-0.6, -0.001, 0.002, 0.005, 0.016, 0.06, 0.2)$.}\label{fig:xi_Planck}
\end{figure}

The RSD correlation function is given by the Fourier transformation of Eq.~\ref{eq:Pkth} as,
\ba\label{eq:xi_eq}
\xi_s(\sigma,\pi)&=&\int \frac{d^3k}{(2\pi)^3} P^{\rm (S)}(k,\mu)e^{i{\bf k}\cdot{\bf s}}\nn\\
&=&\sum_{\ell:{\rm even}}\xi_\ell(s) {\cal P}_\ell(\nu)\,,
\ea
where ${\cal P}_\ell$ are the Legendre polynomials, $\nu=\pi/s$ and $s=(\sigma^2+\pi^2)^{1/2}$. The $\ell$-th moment of correlation function $\xi_\ell(s)$, is defined by,
\ba
\xi_\ell(s)=i^\ell\int\frac{k^2dk}{2\pi^2}\,P^{\rm (S)}_\ell(k)\,j_\ell(ks)\,.
\ea
For the improved model given in Eq.~\ref{eq:Pkth}, the multipole power spectra $P^{\rm (S)}_\ell(k)$ are explicitly given by \cite{2014PhRvD..89j3541S}. The theoretical two--point correlation functions with varying $m_\nu$ are presented in Fig.~\ref{fig:xi_Planck}. The black solid, dash and dotted contours represent $\xi_s(\sigma,\pi)$ with $m_\nu=0\,,0.2\,,0.5\eV$ respectively. The effect of neutrino mass is mainly observed at outer contours. While BAO ring is not altered with varying $m_\nu$, the locations of BAO peak run away from the pivot peak point at $\xi_s(\sigma,\pi)\sim 0.016$.

\subsection{BOSS DR11 catalogue and estimators}

The analysis presented in this work is performed on the 11th data release of the Baryon Oscillation Spectroscopic Survey (BOSS)~\cite{2012AJ....144..144B, 2013AJ....145...10D, 2013AJ....146...32S}, a survey of the Sloan Digital Sky Survey (SDSS)~\cite{2000AJ....120.1579Y, 2006AJ....131.2332G}. In this section we briefly summarise the data we use. However, more detailed information including selection cuts and systematics are explained in~\cite{2014JCAP...12..005S}. Depending on its target, BOSS has the two principal galaxy samples, one is the Constant Stellar Mass Sample (CMASS) and the other is LOWZ. In CMASS \cite{CMASS, 2012MNRAS.424..564R} we use about 690,000 galaxies within the redshift range of $z =0.4-0.7$ over a sky coverage of 8,500 square degrees in the effective volume about 6 Gpc$^3$.

The selected galaxies are observed in (RA, Dec, z) coordinates, and converted into comoving coordinates using a fiducial cosmology. The possible discrepancy between fiducial and true cosmologies leaves an additional geometrical distortion on the anisotropic two--point correlation function, which is dubbed, the Alcock-Paczynski effect (hereafter AP effect)~\cite{1979Natur.281..358A, 2008PhRvD..77l3540P, 2009MNRAS.399.1663G, 2014ApJ...781...96L}. The anisotropic two--point correlation function $\xi(\sigma,\pi)$ is measured using the Landy-Szalzy estimator~\cite{1993ApJ...412...64L}. A random catalogue with about 20 times more number of points than the galaxy data is provided to probe the clustering strength of galaxies relative to the unclustered background. 

The separation distances of $\sigma$ and $\pi$ range from 0$<$ $\sigma$, $\pi$ $<$ 200 Mpc with linear spacing of $\Delta\sigma\,,\Delta\pi=10\mpc$. The covariance matrix between different bins is computed using 600 simulated catalogues mocking the survey geometry and number density~\cite{2013MNRAS.428.1036M}. Further details regarding these simulations including initial conditions and methodology are explained in~\cite{2012MNRAS.426.2719R,2014JCAP...12..005S}. 

This analysis is not valid at smaller scales below $\sigma_{\rm cut}<60\mpc$ and $s_{\rm cut}<75\mpc$. Our understanding of RSD theory is mainly limited by non--linear corrections. The incomplete knowledge of perturbation theory results in unpredictable FoG effect as well. This theoretical limit is fully investigated in our previous work~\cite{2014JCAP...12..005S}, and assumed that the known cut-off scales are weakly dependent on cosmological models.



\subsection{Cosmological model independent constraint on $m_{\nu}$}

\begin{figure*}
\begin{center}
\resizebox{6.5in}{!}{\includegraphics{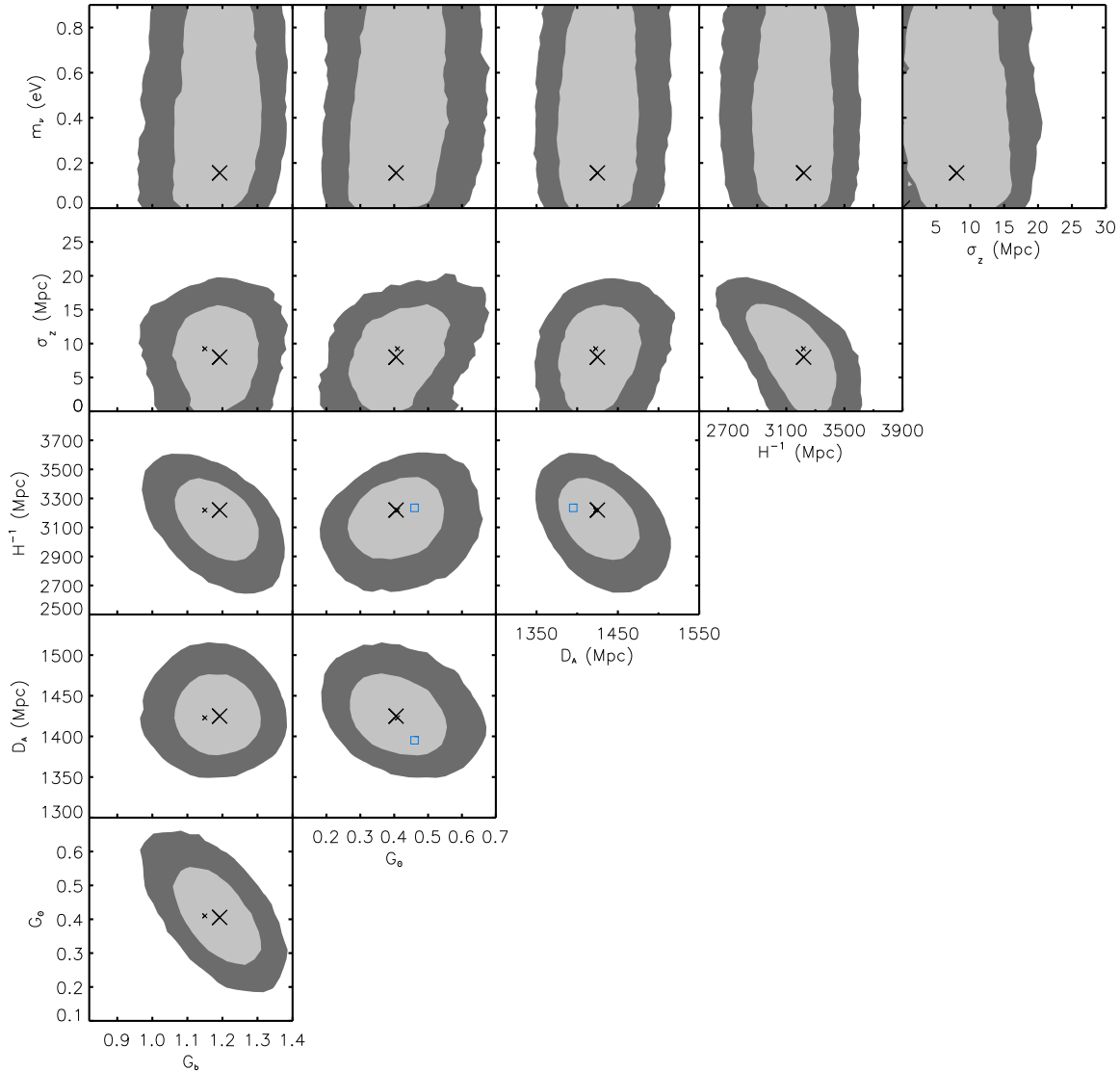}}
\end{center}
\caption{Constraints on observables $(D_A,H^{-1},G_b,G_\Theta,\sigma_z,m_\nu)$ are presented in multi--correlated planes. The blue square points represent the fiducial values of the Planck $\Lambda$CDM concordance model. The black small and large cross points represent the best fit values without and with $m_\nu$ respectively.}\label{fig:result1}
\end{figure*}

In the mild neutrino mass limit of $m_\nu\la 1\eV$, the effect of non--relativistic neutrino is not imprinted on the early broadband shape of the power spectrum, which is solely determined by matter--radiation competition before the last scattering surface. When the neutrino becomes non--relativistic at late time, the neutrino starts to contribute to the clustering, and the early broadband shape becomes alternated. The signature of non--zero $m_\nu$ is probed by analysing the anisotropy of $\xi(\sigma,\pi)$ imprinted by this alternated late time broadband. The neutrino mass is simply added to the parameter space which includes all other cosmological model independent observables such as cosmic distances, growth functions and FoG effect. The new parameter space is given by $(D_A,H^{-1},G_b,G_\Theta,\sigma_z,m_\nu)$. When the broadband shape is invariant, one single theoretical RSD template is sufficient to reproduce $\xi(\sigma,\pi)$. But when $m_\nu$ is added, the multiple theoretical RSD templates are exploited from $m_\nu=0$ to $1\eV$ with $\Delta m_\nu=0.1\eV$. Those templates are interpolated to find the RSD model for each $m_\nu$. 


The constraint on the neutrino mass is presented in Table~\ref{tab:measured}. A small neutrino mass is slightly favoured, but we are not able to determine the lower nor the upper bounds clearly. If no specific dark energy models are given a priori, the neutrino mass is not determined with any precision using BOSS DR11. This poor constraint on $m_\nu$ is well visualised in the five top panels of Fig.~\ref{fig:result1}. The signature of the late time broadband alteration by neutrino damping effect is a very weak signal to be seen from this observation. 

\begin{table}
\begin{center}
\begin{tabular}{lccccc}
\hline
\hline
Parameters & fiducial & without $m_\nu$ & with $m_\nu$\\
\hline
$D_A\,(\mpc)$      & $1395.2$ & $1422.9^{+29.5}_{-32.2}$    & $1424.9^{+33.9}_{-28.8}$ &\\
$H^{-1}\,(\mpc)$   & $3234.69$ & $3218.3^{+203.1}_{-175.5}$  & $3219.5^{+205.4}_{-151.4}$ &\\
$G_b$ &--- & $1.15^{+0.08}_{-0.08}$ & $1.19^{+0.08}_{-0.08}$ &\\
$G_{\Theta}$ &$0.46$ & $0.41^{+0.09}_{-0.09}$ & $0.41^{+0.08}_{-0.10}$ &\\
$\sigma_z\, (\mpc)$ &---&$9.2^{+5.4}_{-5.7}$&$8.0^{+4.4}_{-5.1}$&\\
$m_\nu\, (\text{eV})$ &---&---&$0.16^{(+0.59)}$&\\
\hline
\hline
\end{tabular}
\end{center}
\caption{Constraints on values of $D_A$, $H^{-1}$, $G_b$, $G_{\Theta}$, $\sigma_z$, and $m_\nu$ with their 68\% CL uncertainties and compare with the previous study without massive neutrinos \cite{2014JCAP...12..005S}. Note that the reported uncertainty with the parenthesis is calculated with the assumption that the probability distribution is Gaussian.
}\label{tab:measured}
\end{table}

\begin{figure*}
\begin{center}
\resizebox{3.2in}{!}{\includegraphics{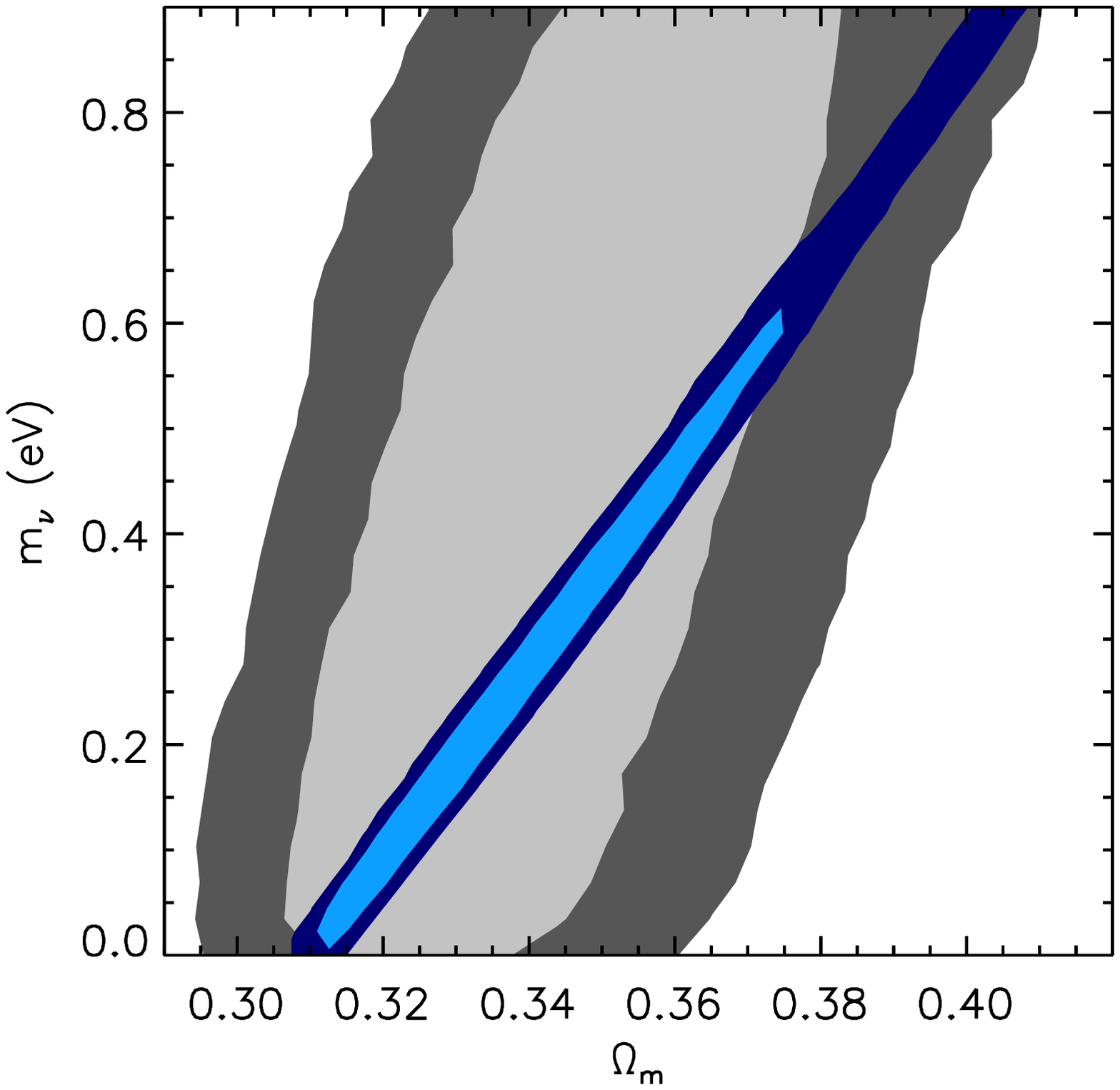}}
\resizebox{2.9in}{!}{\includegraphics{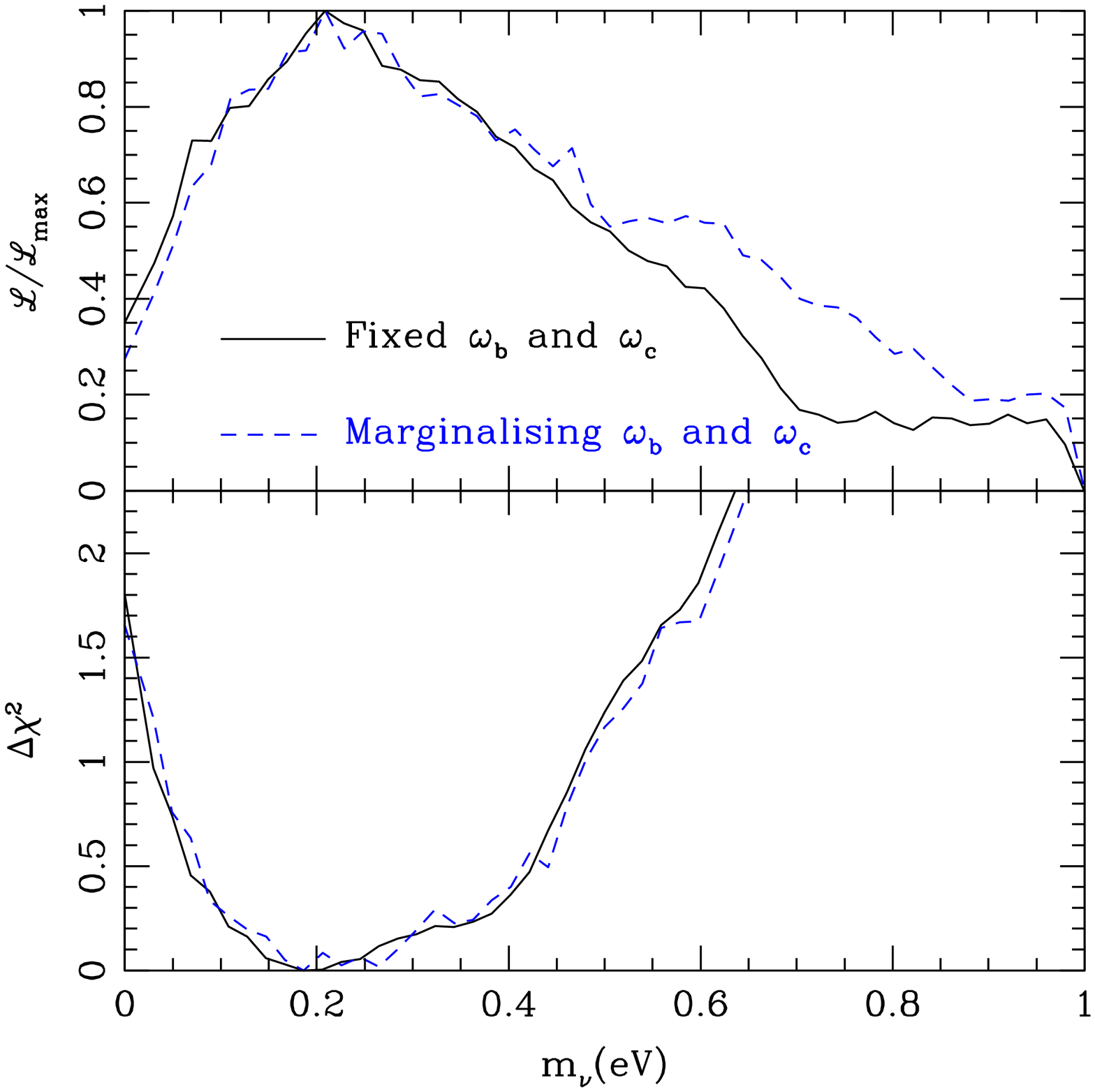}}
\end{center}
\caption{{\it (Left panel)} The 2D contour plot of $(\Omega_m,m_\nu)$ is presented. The black and blue contours represent the constraints using BOSS DR11 only and the combination of BOSS DR11 and Planck distance measures. Inner and outer contours present 68\% and 95\% confidence levels. Here both $\omega_b$ and $\omega_c$ are fixed. {\it (Right panel)} We present the likelihood function and $\Delta\chi^2$ in the top and bottom panels. The black solid and blue dash curves represent the cases with the fixed and marginalised $\omega_b$ and $\omega_c$ respectively.}\label{fig:result2}
\end{figure*}

However, we are more interested in the consistency checks for all other observables of $(D_A,H^{-1},G_b,G_\Theta,\sigma_z)$ which were reported without $m_\nu$ in our previous works \cite{2014JCAP...12..005S}. The one--point FoG effect is known to be well represented by Gaussian functional form, but the first order contribution of it is dominating, which causes a significant degeneracy with the coherent motion constraint. This random velocity effect is represented by a single parameter $\sigma_z$, which is presented in the second panels of Fig.~\ref{fig:result1}. The small and large cross points represent the best fitting values without and with $m_\nu$ respectively, and we find that the contamination due to random velocity effect is not much altered with including small neutrino mass. 

The distance measures of $D_A$ and $H^{-1}$ are least impacted by neutrino mass. The measured $D_A$ and $H^{-1}$ without and with $m_\nu$ are nearly equivalent to each other. The size of the contours do not change significantly with the additional degree of freedom $m_\nu$ either. The physical scales associated with the primordial BAO features imprinted on large scale structure remains unaltered, and it claims that both distance measures $D_A$ and $H^{-1}$ are mainly determined by BAO peaks, rather than the overall shape of the spectrum. 

Contrary to the other parameters, the influence of non--trivial $m_\nu$ on $G_b$ is observed as an increase of around 5\%. This occurs at the best fit $m_\nu=0.2\eV$ which is away from the massless limit. The combined effect of the coherent growth and $m_\nu$ damping effect is determined at the weight centre in the effective $k$ range. The late time broadband shape is observed to be pivoting this weight $k$ centre around $k=0.05\ompc$. When $m_\nu=0.2\eV$ is favoured by the fitting, the shape of the spectrum is pivoted down toward the smaller scales, and boost the large scale part. As the coherent growth function is defined at $k\rightarrow 0$, it causes the 5\% increment for the measured $G_b$. This correlation between $m_\nu$ and $G_b$ is not clearly visible from the contour, because $m_\nu$ is poorly determined. The similar effect is not seen from $G_\Theta$, as $G_\Theta$ is not determined precisely unlike $G_b$.

\subsection{Constraint on $m_{\nu}$ with the standard model prior}

\begin{table}
\begin{center}
\begin{tabular}{lccccc}
\hline
\hline
Parameters & B & BP  & BP (marginalising $\omega_b, \omega_c$)\\
\hline
$\Omega_m (0.32)$      &  $0.35^{+0.02}_{-0.02}$ & $0.34^{+0.03}_{-0.02}$    & $0.34^{+0.03}_{-0.02}$ &\\
$m_\nu(0\eV)$   & $0.28^{(+0.62)}$ & $0.19^{+0.28}_{-0.17}$  & $0.18^{+0.30}_{-0.15}$ &\\
$b$ &$1.98^{+0.11}_{-0.11}$ & $1.96^{+0.09}_{-0.10}$ & $1.95^{+0.09}_{-0.10}$ &\\
$\sigma_z\, (\mpc)$ &$8.0^{+3.0}_{-3.7}$&$8.1^{+3.0}_{-3.5}$&$8.0^{+3.1}_{-3.7}$&\\
\hline
\hline
\end{tabular}
\end{center}
\caption{Constraints on values of $\Omega_m$, $b$, $\sigma_z$ and $m_\nu$ with their 68\% CL uncertainties and compare with the previous study without massive neutrinos \cite{2014JCAP...12..005S}. In the first low, ``B" and ``P" represent the constraints using BOSS DR11 and Planck experiments respectively. Again, the reported uncertainty with the parenthesis is calculated with the assumption that the probability distribution is Gaussian.}\label{tab:measured2}
\end{table}

The cosmic acceleration has been confirmed by many efforts since the first discovery in 1998 \cite{1999ApJ...517..565P, 1998AJ....116.1009R}, suggesting new physics to modify the knowledge of materials or gravity. The unknown materials such as dark energy can be a solution to expel the cosmic expansion, or our incomplete understanding of gravitational physics at large scale is a cause of acceleration. Among those all exotic explanations, the cosmological constant can be a most compromising solution between the standard model of particle physics and the cosmological observations. If we work under the standard model frame, the neutrino mass can be considered within the $\Lambda$CDM model, in which 95\% dark materials are possibly explained without significant modification on the known physics. The massive neutrino is a necessary element of standard model which is theoretically predicted, and $\Delta m^{2}_{\nu, \mu, e}$ is observationally confirmed to be non-zero. Thus the $\Lambda$CDM model with non--trivial $m_{\nu}$ is the best theoretical model to be consistent with both particle and cosmological experiments. 

The cosmological model dependent parameter space is $(\Omega_m, m_\nu,b,\sigma_z)$, where both $\omega_b$ and $\omega_c$ are given or marginalised under the CMB experiment constraints. We adopt the tight constraint on primordial parameters from CMB experiments, and we drop out all other nuisance parameters such as the optical depth $\tau$. Then there are two major cosmological parameters of $(\Omega_m, m_\nu)$, and two systematical parameters of $(b,\sigma_z)$. Theoretical RSD templates are provided in two dimensional space of cosmological parameter $(\Omega_m, m_\nu)$. 

The constraint on $m_{\nu}$ using BOSS DR11 alone is presented in column ``B" as $0.28^{(+0.62)}\eV$. The lower bound is not determined, and the upper bound is given with the assumption of Gaussianity of the probability distribution. Then the upper bound of $m_{\nu}$ is $m_{\nu}\la 0.9\eV$. Considering the small $m_{\nu}\la 1\eV$ limit, we do not think that the upper bound of $m_{\nu}$ is determined. The measured $\Omega_m$ becomes bigger than the value from the Planck $\Lambda$CDM concordance model, which is caused by bigger mean $m_{\nu}$. The measured galaxy bias is consistent with the expected CMASS galaxy bias.

The benefit of cosmological model dependent parameterisation is to combine other cosmological tests to break degeneracy. The CMB measures the angular extension of the sound horizon at last scattering surface, which is given by $\theta_*=1.04\pm0.00047$ \cite{2015arXiv150201589P}, of which the constraint is denoted as ``P" in Table~\ref{tab:measured2}. The matter content of $\Omega_m$ is precisely determined for $\Lambda$CDM cosmology with massless neutrino using $\theta_*$ constraint alone. But $\Omega_m$ becomes undetermined with varying $m_\nu$. Both are tightly correlated to each other on $(\Omega_m, m_\nu)$ plane. This indefinite degeneracy is broken by the combination of BOSS and $\theta_*$ constraints, which is presented in the left panel of Fig.~\ref{fig:result2}. The black and blue contours represent the constraints on $(\Omega_m, m_\nu)$ using BOSS DR11 alone and the combined BOSS DR11 and  $\theta_*$ measurement. The upper bound becomes tighter, and the lower bound of $m_\nu$ becomes visible.

In Table~\ref{tab:measured2}, $m_\nu$ is reported to be $m_\nu=0.19^{+0.28}_{-0.17}\eV$ when $\omega_b$ and $\omega_c$ are fixed. The error is estimated with the assumption of Gaussian distribution again. Both likelihood function and $\Delta \chi^2$ are presented in the right panel of Fig.~\ref{fig:result2}, where $\Delta \chi^2=128.0$ at $m_\nu=0.2\eV$ and $\Delta \chi^2=129.8$ at $m_\nu=0.0\eV$. It suggests that $m_\nu$ is not massless. Planck probes $\omega_b$ and $\omega_c$ tightly, so that the variation of both provides the minimal difference in the fitting results using BOSS DR11 alone. But the influence on cosmological constraints using $\theta_*$ measurement is not minimal. We compare the $m_\nu$ constraints without and with $\omega_b$ and $\omega_c$ marginalisation as black solid and blue dash curves in the right panel of Fig.~\ref{fig:result2}, respectively. There is no significant difference observed. The measured $m_\nu$ is given in Table~\ref{tab:measured2} as $m_\nu=0.18^{+0.30}_{-0.15}\eV$.

The estimated Hubble constant $H_0$ becomes $H_0=65\pm1.3 {\rm \,km/s/Mpc}$ with $\Lambda$CDM cosmology with massive neutrino. The recent direct measurement of $H_0$ is $73\pm2 {\rm \,km/s/Mpc}$ \cite{2016arXiv160401424R}, which is compared with the Planck estimation when $\Lambda$CDM cosmology with massless neutrino is assumed, $H_0=67\pm0.96 {\rm \,km/s/Mpc}$. Note that the non--trivial neutrino mass does not resolve the discrepancy, but makes it worse.

\section{Conclusion}


The early broadband shape of the power spectrum is altered by the damping effect due to the massive neutrino with $m_\nu\la 1\eV$ at later epoch. The observed galaxy clustering in redshift space can be exploited to observe this signature imprinted by non--trivial neutrino mass. We provide the methodology to detect $m_\nu$ with probing this altered broadband shape of power spectrum. This damping effect can be disclosed simultaneously with distance measures and coherent growth factors which are dependent on the unknown dark energy models. When the origins of cosmic acceleration is completely unknown, i.e. neither distance measures nor coherent growth functions are known, the neutrino mass is not measured in precision using BOSS DR11 catalogues. But we find that both distance measures and growth functions are minimally affected by $m_\nu$ marginalisation. It justifies our previous results without including $m_\nu$ damping effect.

Although there is no confirming evidence of cosmological constant, the most reliable and simplest theoretical model to explain the cosmic acceleration is $\Lambda$CDM model. We continue our test with a theoretical prior that the cosmic expansion is accelerated by the existence of cosmological constant. Then all distance measures and growth functions vary coherently with one single cosmological paramter $\Omega_m$. Despite this reduction of parameter space, the neutrino mass is not clearly distinguishable from the massless limit, and the upper bound is poorly set by $m_\nu\la 0.8\eV$. However, the hidden $m_\nu$ can be revealed by ``smoking gun" from the combination of CMB distance measure at the last scattering surface. The angular extension of sound horizon constrain the $\Omega_m$ and $m_\nu$ correlation pattern. The likelihood plane of ($\Omega_m,m_\nu$) probed by BOSS DR11 is sharply sliced by the determined trajectory of $\Omega_m$ and $m_\nu$ correlation by CMB distance measure, presented in Fig.~\ref{fig:result2}. This combination excavates the hidden neutrino mass as $m_{\nu} = 0.19^{ +0.28}_{ -0.17}\eV$. The statistical error is reported with assuming Gaussianity of probability distribution. Although this assumption is not made, both likelihood function and $\Delta \chi^2$ results in Fig.~\ref{fig:result2} confirms the fact that $m_\nu$ is probed to be different from massless limit, which was also reported in \cite{2014MNRAS.444.3501B} using different approach.

While we focus on probing neutrino mass through galaxy clustering observed by BOSS DR11 with the assistance of CMB distance measures, CMB also contains an alternative clustering information imprinted by lensing effect. The signature of neutrino mass is also preserved in the distorted CMB anisotropy maps at smaller scales. Thus, instead of taking only CMB distance measure constraint,  the full combination between BOSS DR11 and CMB measurements is an interesting topic to be investigated. Considering the targeted volume of BOSS DR11, the observed clustering by BOSS DR11 affects the measured CMB lensing as well about 20\%. We would like to address the full covariance approach in the following work, we plan to probe $m_\nu$ using not only galaxy clustering but also CMB lensing effect. 


The measured constraints on $m_\nu$ presented in this manuscript is trustable with low resolution catalogues provided by BOSS DR11. But we do not think that the same theoretical templates are applicable for the future precision experiments. The same methodology can be used with more rigorously developed RSD theoretical templates. We work towards to improve those using the detailed perturbation theory and the believable neutrino simulations in near future.


\section*{Acknowledgements}
We thank Eric Linder, Cris Sabiu and Yi Zheng for helpful comments, and thank Atsushi Taruya for a piece of advice on RSD templates. Numerical calculations were performed by using a high performance computing cluster in the Korea Astronomy and Space Science Institute. Funding for SDSS-III has been provided by the Alfred P. Sloan Foundation, the Participating Institutions, the National Science Foundation, and the U.S. Department of Energy Office of Science. The SDSS-III web site is http://www.sdss3.org/. SDSS-III is managed by the Astrophysical Research Consortium for the Participating Institutions of the SDSS-III Collaboration including the University of Arizona, the Brazilian Participation Group, Brookhaven National Laboratory, Carnegie Mellon University, University of Florida, the French Participation Group, the German Participation Group, Harvard University, the Instituto de Astrofisica de Canarias, the Michigan State/Notre Dame/JINA Participation Group, Johns Hopkins University, Lawrence Berkeley National Laboratory, Max Planck Institute for Astrophysics, Max Planck Institute for Extraterrestrial Physics, New Mexico State University, New York University, Ohio State University, Pennsylvania State University, University of Portsmouth, Princeton University, the Spanish Participation Group, University of Tokyo, University of Utah, Vanderbilt University, University of Virginia, University of Washington, and Yale University.





\providecommand{\href}[2]{#2}\begingroup\raggedright\endgroup

\end{document}